# Journal Portfolio Analysis for Countries, Cities, and Organizations: Maps and Comparisons

*Journal of the Association for Information Science and Technology* (in press)

Loet Leydesdorff,*[a] Gaston Heimeriks,[b] and Daniele Rotolo[c]

**Abstract**

Using Web-of-Science data, portfolio analysis in terms of journal coverage can be projected on a base map for units of analysis such as countries, cities, universities, and firms. The units of analysis under study can be compared statistically across the 10,000+ journals. The interdisciplinarity of the portfolios is measured using Rao-Stirling diversity or Zhang *et al.*'s (in press) improved measure $^2D^3$. At the country level we find regional differentiation (e.g., Latin-American or Asian countries), but also a major divide between advanced and less-developed countries. Israel and Israeli cities outperform other nations and cities in terms of diversity. Universities appear to be specifically related to firms when a number of these units are exploratively compared. The instrument is relatively simple and straightforward, and one can generalize the application to any document set retrieved from the Web-of-Science (WoS). Further instruction is provided online at http://www.leydesdorff.net/portfolio.

**Keywords**: map, journal, portfolio, firm, university, city, country

[a] * corresponding author; Amsterdam School of Communication Research (ASCoR), University of Amsterdam PO Box 15793, 1001 NG Amsterdam, The Netherlands; email: loet@leydesdorff.net
[b] Department of Innovation Studies, Faculty of Geosciences, Utrecht University, Heidelberglaan 2, NL-3584 CS Utrecht, The Netherlands; tel.: +31-30-253 7802; email: gheimeriks@gmail.com .
[c] SPRU – Science Policy Research Unit, University of Sussex, Brighton, BN1 9SL, United Kingdom; D.Rotolo@sussex.ac.uk

## 1. Introduction

Like other forms of portfolio management (see for a recent literature review Wallace & Rafols, in press; cf. Rafols *et al.*, 2010; Zhang *et al.*, 2011), portfolio analysis in terms of journals may provide insights into the specialization of countries, cities, or knowledge-producing organizations such as universities and firms. Analytically, the matrix of journals *versus* countries has been basic to evaluative bibliometrics (Narin, 1976; Small & Garfield, 1985). In this brief communication, we introduce a generalized instrument to generate such a matrix for the purpose of mapping and analyzing portfolios using tools available online at the Web-of-Science (WoS) and http://www.leydesdorff.net/portfolio.

The base map onto which the portfolios can be overlaid was provided by Leydesdorff, Rafols, & Chen (2013). Portfolios can be disaggregated at the city-level, the level of organizations, or— more generally—any document set retrieved from WoS. In addition to the visuals (using VOSviewer; Van Eck & Waltman, 2010), the data can be analyzed statistically using the matrix generated in each analysis in formats compatible to SPSS and Pajek/UCINET. Analytically, this further extension enables the user to compare among units (e.g., firms), whereas the visual maps provide an overview of the results.

## 2. Methods and materials

First, the user is invited to identify a document set by using the "Advanced Search" interface of WoS. The identified documents can be examined online using the analytical interface of WoS, namely "Analyze Results". In order to map these documents across journals, in this interface, the user chooses to rank the output in terms of "*source titles*," then ticks "*all data rows*," and saves the file "*analyze.txt*". This file contains the list of journal names where the identified documents were published and the numbers of documents for each journal name.

On February 12, 2015, for example, we searched for all documents involving at least one organization based in the Netherlands and published in the year 2013. The following search string was used in the advanced interface of WOS: "*cu=Netherlands and py=2013*". This recalled 49,000 documents listed in 4,632 of the 10,542 source titles/journal names in the Science Citation Index and Social Science Citation Index of WoS. We use 2013-data throughout this study because, at the date of this research, the indexing of documents published in the year 2014 was not yet complete.

The file *analyze.txt* should be renamed. In this case, we renamed the file "*nl.txt*". The routine *portfolio.exe* prompts the user for this file name and then generates a file "*nl.vos"* that can be opened directly in VOSviewer. Figure 1 depicts the map generated from the *nl.txt* file.[1] Figure 1 shows that the Netherlands has considerable coverage in most journals contained on this map. However, a cluster of journals without coloring at the bottom of the map can be identified as journals interfacing psychology and psycho-analysis. The interactive map (in VOSViewer) enables the user to explore the associated journal names in considerable detail.

---

[1] The coloring of the map is based on the community-finding algorithm in VOSviewer (Leydesdorff, Rafols, & Chen, 2013; Waltman, Van Eck, & Noyons, 2010).

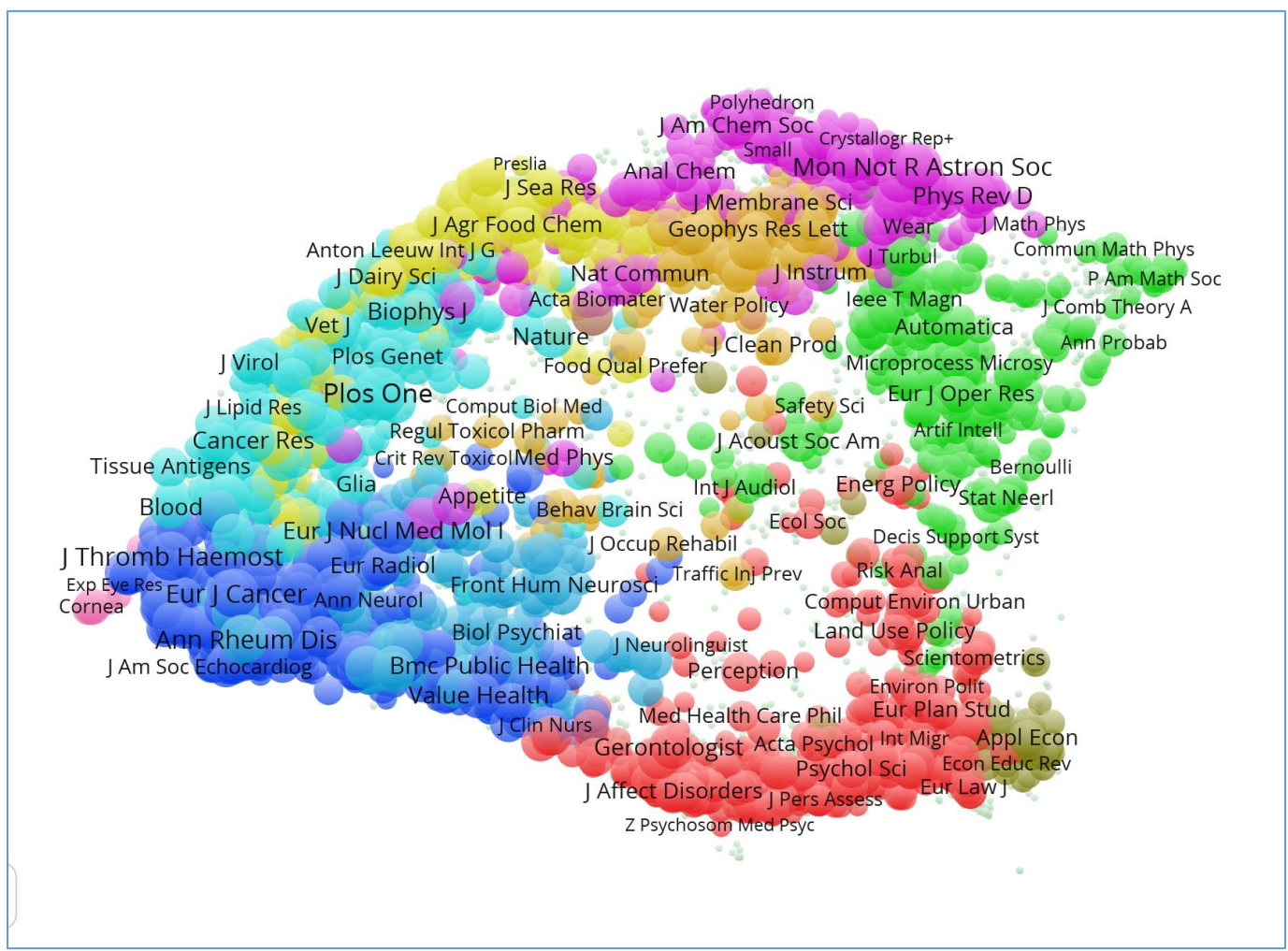

**Figure 1**: Journal Portfolio Map for the Netherlands in 2013. Source: Web-of-Science. (This map can be web-started at http://www.vosviewer.com/vosviewer.php?map=http://www.leydesdorff.net/portfolio/nl.vos.)

The routine *portfolio.exe* also generates the Rao-Stirling diversity value (Rao, 1982; Stirling, 2007) and the modification of this measure ($^2D^3$) recently proposed by Zhang *et al*. (in press). These measures are reported in the file "*rao.dbf*". Diversity can be considered as a measure of the interdisciplinarity of the portfolios under study (Stirling, 2007; cf. Rafols & Meyer, 2010).

The vector containing the information of the number of documents for each of the 10,000+ journals is saved as an additional column in the file *"matrix.dbf"*. This file enables the user to compare vectors for different units of analysis (e.g., in terms of their cosine-normalized similarities).[2] After finishing the analysis for a set of units to be compared, one can run *"mtrx2cos.exe"* that generates the files *"cosine.net"* and *"coocc.dat"* in the Pajek and UCINET formats, respectively, for the purpose of network visualization and analysis. (After deleting the files *"matrix.dbf"* and *"rao.dbf"*, these files are regenerated from scratch for a new round of analyses.)

Rao-Stirling diversity is defined as follows (Rao, 1982; Stirling, 2007):

$$\Delta = \sum_{ij} p_i p_j d_{ij} \qquad (1)$$

where $d_{ij}$ is a disparity measure between two classes *i* and *j*—the categories are in this case journals—and $p_i$ is the proportion of elements assigned to each class *i*. As the disparity measure, we use the distances on the map (Leydesdorff, Rafols, & Chen, 2013).[3] The coordinates for each journal on the map are provided in a companion file *"citing.dbf"* that can also be obtained from the website.

---

[2] Each vector is stored in the matrix as a variable with the original file name as a label, in this case "NL". For this reason, the name of the original file (i.e. *nl.txt*) name should not contain more than ten characters.
[3] Computation of (1 – cosine) values between each two journal points can become too intensive for interactive usage.

Zhang *et al.* (in press) argues that $^2D^S$ provides a true diversity measure that outperforms Rao-Stirling diversity ($\Delta$) because $^2D^S = 2.0$ is twice as diverse as $^2D^S = 1.0$. In Eq. 6, these authors formulate:

$$^2D^S = 1/(1 - \Delta) \qquad (2)$$

where $\Delta$ is the Rao-Stirling diversity. This improved measure varies from 1 to $\infty$ when $\Delta$ varies from 0 to 1. The transformation is monotonic and the value of $^2D^S$ follows directly from that of the Rao-Stirling diversity using Eq. 2. Both measures are provided for each case in the file "*rao.dbf*". Note that these are diversity measures of each portfolio in terms of the journal composition.

## 3. Results

*3.1. Portfolio analysis at the country-level*

To perform the portfolio analysis at the country-level, we considered the list of 34 OECD member states plus the seven affiliated member economies (i.e., Argentina, China, Romania, Russia, Singapore, South Africa, and Taiwan), and the two other BRICS countries (i.e., Brazil and India). This sample of 43 nations covers 1,753,243 documents, that is 89.4% of the total of 1,963,753 documents indexed in WoS for the publication year 2013, as of the date of the download (21 January 2015).[4]

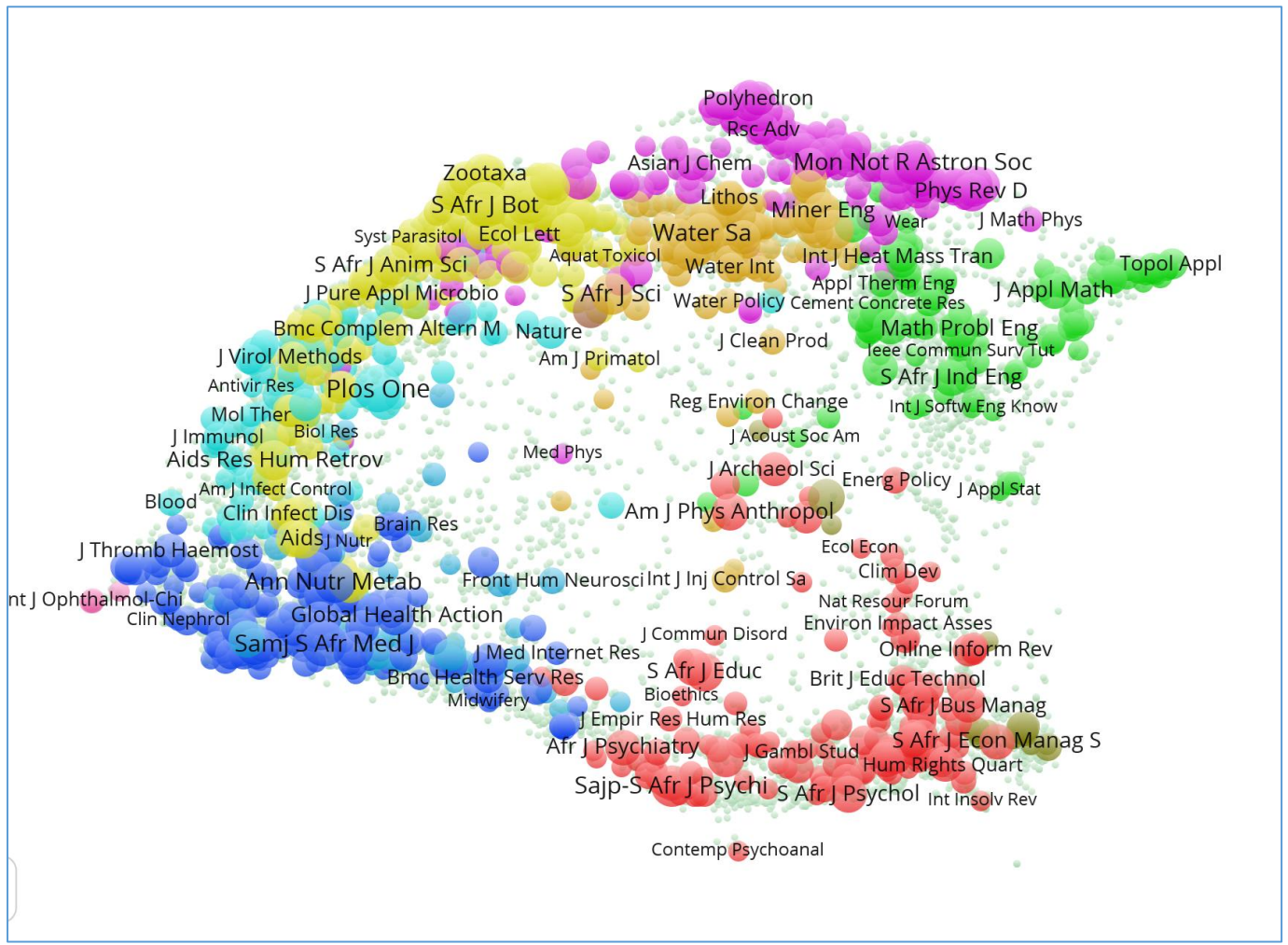

**Figure 2**: Journal Portfolio Map for the South Africa in 2013. Source: Web-of-Science. (This map can be web-started at
http://www.vosviewer.com/vosviewer.php?map=http://www.leydesdorff.net/portfolio/sa.vos.)

Figure 2 shows the portfolio for South Africa, analogous to Figure 1 for the Netherlands. Maps for the other nations included can be web-started using their respective two-character abbreviations instead of "sa" in the string provided in the legend of Figure 2.[5] South Africa, for example, has a relatively weak portfolio in computer sciences and statistics (at the right side of the figure). For most OECD countries, however, the coverage is almost complete (as in the case of the Netherlands).

---

[4] Because this is whole-number counting, the number of records with addresses in these countries aggregates to 2,226,237. Internationally co-authored publications are counted with full counts at the address level.

*3.2.    Comparing portfolios among nations*

After cosine-normalization of the vectors using "*mtrx2cos.exe*", Figure 3 shows a clear divide between the more advanced nations in the world of scientific publishing (red) versus the other nations, including the Mediterranean and Latin-American ones (grey). The clustering and coloring is performed by using VOSViewer, but the results are consistent with those found using other community-finding algorithms (e.g., Blondel *et al*., 2008).

---

[5] This country code table of the ISO (*International Organization for Standardization*) is available at http://www.worldatlas.com/aatlas/ctycodes.htm and http://patft.uspto.gov/netahtml/PTO/help/helpctry.htm

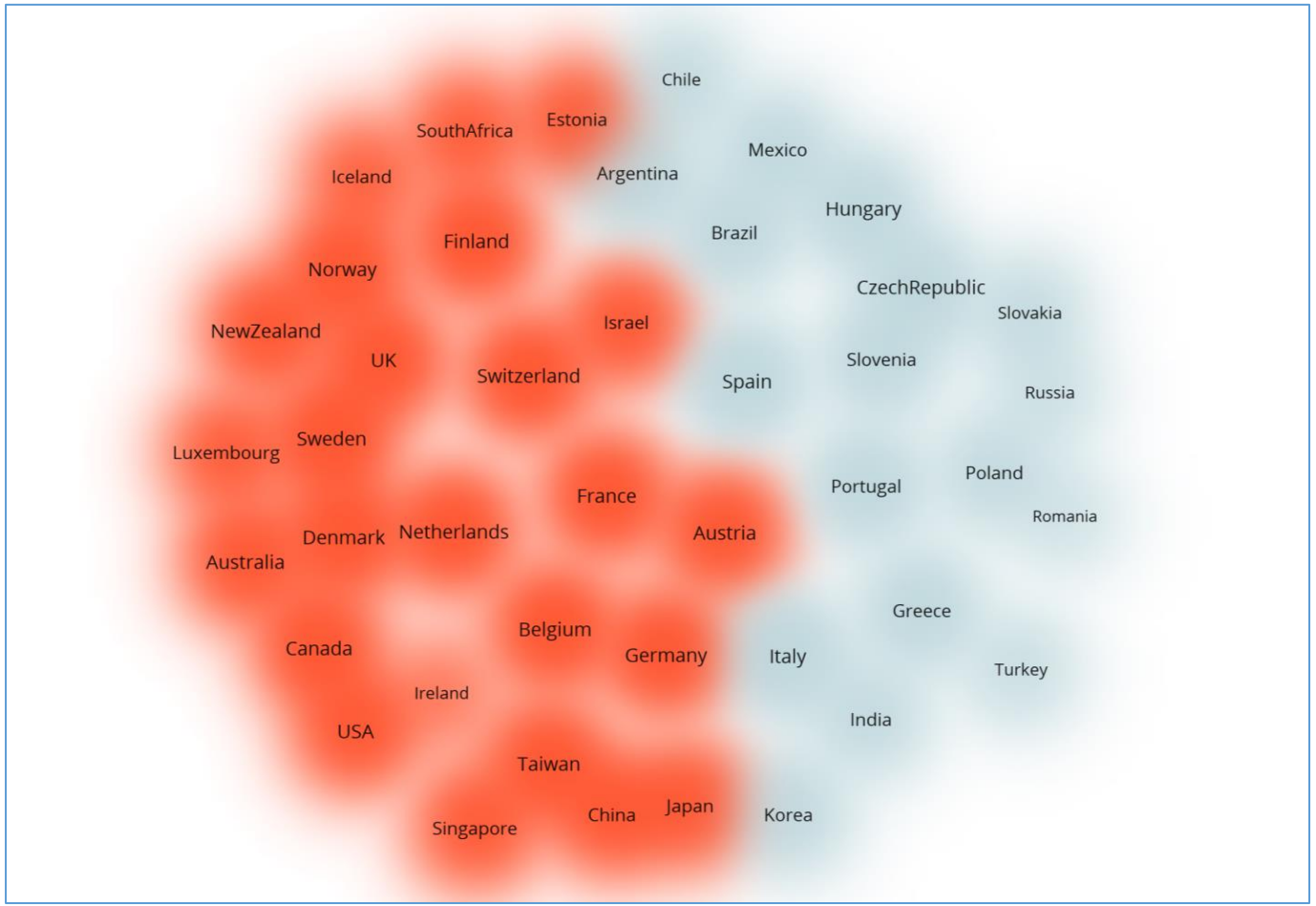

**Figure 3**: Publication patterns compared among 43 nations; based on cosine values classified and mapped using VOSviewer. (This map can be web-started at http://www.vosviewer.com/vosviewer.php?map=http://www.leydesdorff.net/portfolio/cos.vos&label_size=1.20&view=3&white_background&white_background .)

Using factor analysis in SPSS (v.21) with the countries as variables, a five-factor solution (Varimax rotated) sorts the Eastern European countries including Russia to a second group, the Asian countries into a third, the Latin American ones into a fourth, and Greece and Turkey into a fifth group. South Africa is classified with the Latin American countries, but with interfactorial complexity to the first factor that represents the advanced nations. (Similarly, Argentine, China, Taiwan, and Singapore exhibit a second loading on this first factor.)

This can also be made visible using the affiliations matrix of co-occurrences. Figure 4 shows a first divide between the Anglo-Saxon/Scandinavian world with some other nations *versus* the remainder of the continental EU. The latter, including Canada, is now the strongest group because of transnational within-EU collaborations. Belgium, Switzerland, and Canada show separate profiles—as expected because of their bi- and multi-lingual cultures. Israel is also a separate group for reasons that we shall discuss further below.

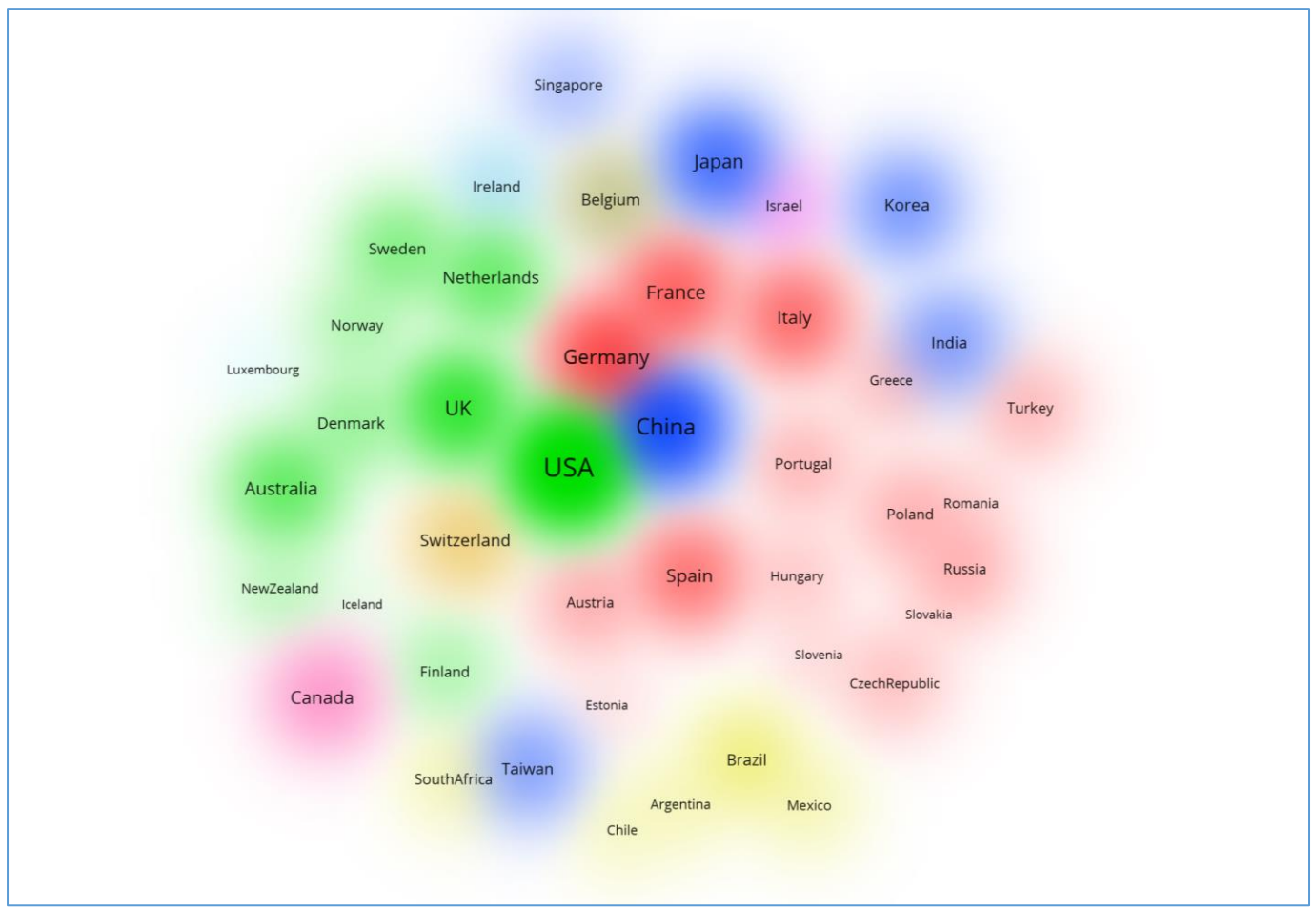

**Figure 4**: Publication patterns compared among 43 nations; based on the affiliations matrix classified and mapped using VOSviewer. (This map can be web-started at http://www.vosviewer.com/vosviewer.php?map=http://www.leydesdorff.net/portfolio/coocc.vos&network=http://www.leydesdorff.net/portfolio/netw_coocc.vos&label_size=1.20&n_links=1000&view=3&white_background .)

In summary, these 43 nations can be subdivided into regionally relevant categories such as the Asian nations, depending on the number of components distinguished. In addition to the regional divisions, there is a major divide between advanced and less-advanced nations. The profiles of Japan, China, Singapore, and Taiwan, for example, are classified in the first category; but Korea and India are not. These results provide us with some confidence that the instrument can also be used for units of analysis other than nations, such as cities and organizations, and can provide interesting insights.

*3.3.    Portfolio analysis at the city-level*

Cities can be expected to entertain different portfolios both in terms of their sizes and given the differences among national cultures. Metropolitan cities with multiple universities, for example, will have portfolios different from small towns with a technical university. There are many cities in the world, and many different rankings, such as for "global cities," "innovative cities," etc., are available both in the literature and online (e.g., Matthiesen, Schwarz & Find, 2010; Van Noorden, 2010).

Given the explorative nature of this research, we selected four cities in each of five different countries about which we have some common knowledge so that we might conjecture to have sufficient variety in different dimensions. The five countries under study are: China, France, Israel, the Netherlands, and the USA. The cities are listed in Table 1. We applied again the *portfolio.exe* routine to sets of documents associated with each of these cities.

**Table 1**: Twenty cities in five countries.

| Country | Cities |
|---|---|
| China | Beijing, Shanghai, Nanjing, Dalian |
| France | Paris, Marseille, Grenoble, Toulouse |
| Israel | Jerusalem, Tel Aviv, Haifa, Beer Sheva |
| Netherlands | Amsterdam, Rotterdam, Eindhoven, Wageningen |
| USA | Boston, Atlanta, Berkeley, Boulder |

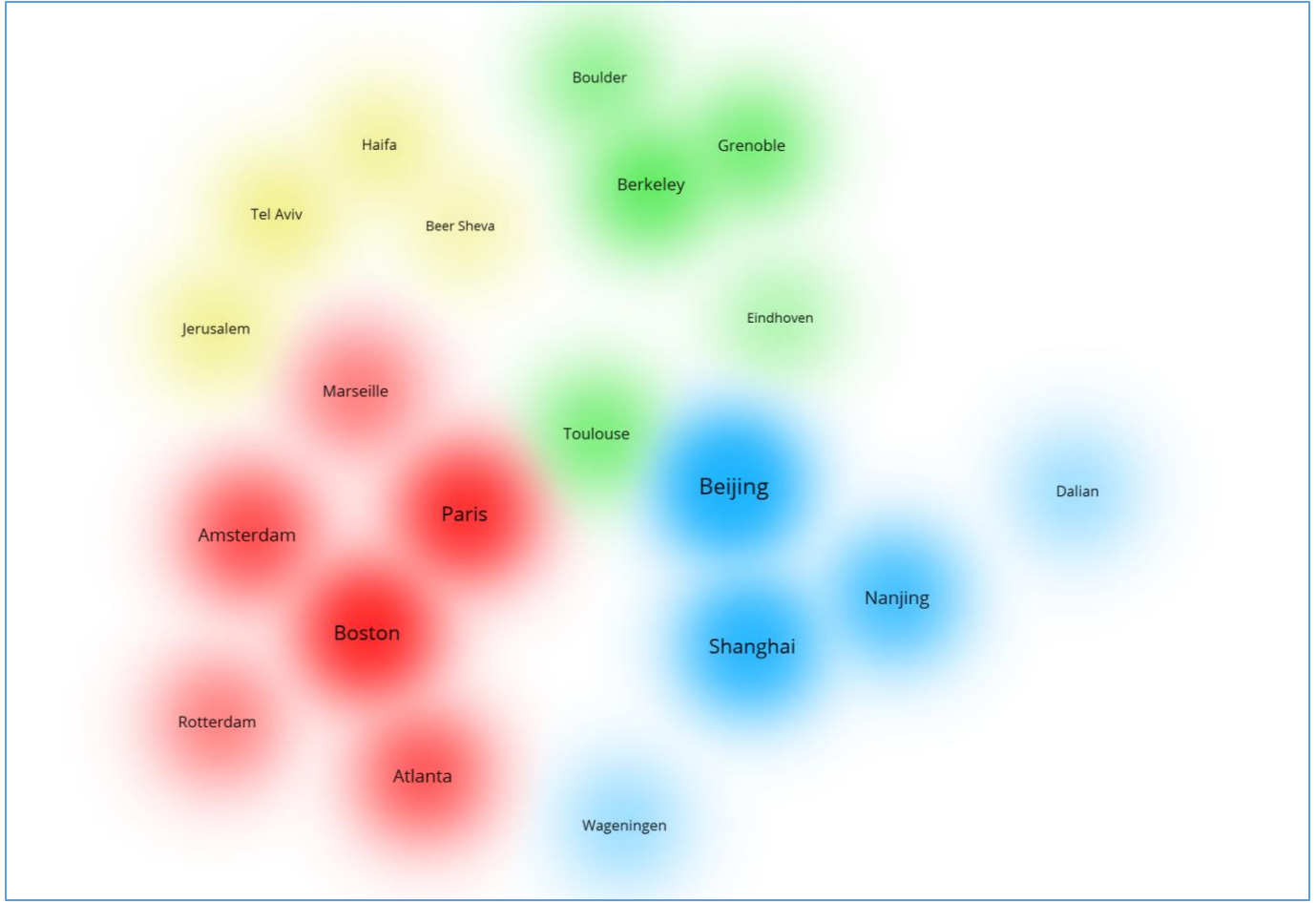

**Figure 5**: Publication patterns compared among 20 cities; based on the affiliations matrix classified and mapped using VOSviewer. (This map can be web-started at http://www.vosviewer.com/vosviewer.php?map=http://www.leydesdorff.net/portfolio/aff_city.map&network=http://www.leydesdorff.net/portfolio/aff_city.net&label_size=1.40&n_links=1000&view=3&white_background .)

Figure 5 shows first that the Israeli cities and universities are grouped separately. The Chinese group is joined by the Dutch city of Wageningen. Wageningen is a small town housing an agricultural university. The other two groups are mixtures of European and American cities. The division, in our opinion, distinguishes cities with city-universities from smaller cities with specific capacities. When the vectors are cosine-normalized, the Israeli cities are part of the latter (green-colored) group, and Toulouse and Eindhoven are drawn into the (red-colored) group of city-universities.

One should note that the level of precision obtained from searching with the city names is not controlled using WoS. WoS uses the address information provided by the authors in the bylines. Many cities are administratively underbounded (e.g., Amsterdam, Rotterdam) and may have universities in suburbs, whereas other cities are overbounded (e.g., Paris). In the USA, Core Based Statistical Areas (CBSA) are defined by the US Office of Management and Budget (OMB). A CBSA is a group of adjacent areas that are socioeconomically close to an urban center. However, series of attempts at constructing a European counterpart to the metropolitan region concept of the US are still short of results, which could be used for the purpose of comparing the scientific base of large cities (Grossetti *et al*., 2013, 2014).

*3.4.     Portfolio analysis at the organization-level: universities and industries*

The choice of organizations is even more difficult to justify than the choice of cities. For the sake of comparability, we performed the analysis on a sample of organizations used in previous studies (Rafols *et al*., 2010; Leydesdorff, Rafols & Chen, 2013) and added to this sample the following companies: Google, Samsung, and Philips. The list of organizations is reported in Table 2.

**Table 2**: Ten organizations mapped and compared.

| Universities | Industries (Rafols *et al.*, 2010) | Industries added |
|---|---|---|
| University of Amsterdam | Pfizer | Google Inc. |
| Georgia Inst. of Technology | Nestlé SA | Samsung |
| London School of Economics | Unilever | Philips |
| | Shell | |

Publications of organizations can be retrieved at WoS using an index of consolidated names. Using, for example, "OG=(Georgia Institute of Technology)" 3,504 records were retrieved with publication year 2013. Extension with "Georgia Tech" provided another 19 records. Whereas these names are reasonably reliable in the case of universities, one is advised to use the common company names in the case of enterprises. The consolidated name "Royal Dutch Shell," for example, did not provide any retrieval for 2013, but 179 publications could be found using "Shell" as the search term (including such names as "Shell Canada Ltd.").

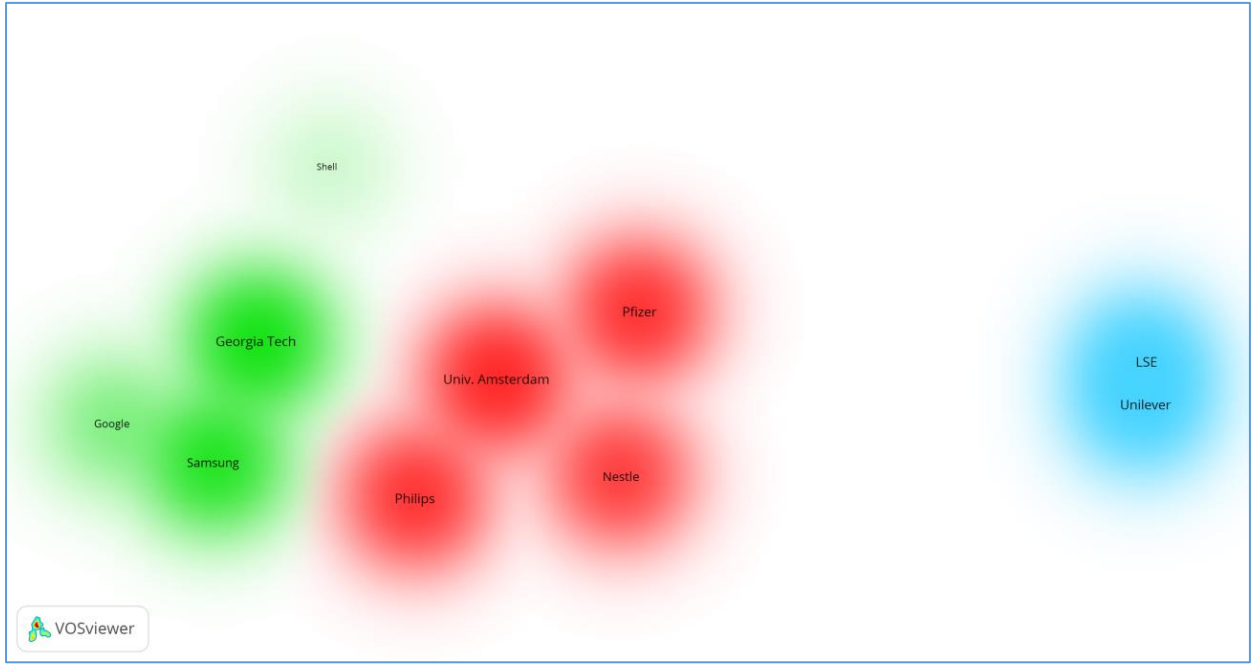

**Figure 6**: Cosine-normalized profiles of three universities and seven firms. (This map can be web-started at http://www.vosviewer.com/vosviewer.php?map=http://www.leydesdorff.net/portfolio/cos_univ.txt&label_size=1.40&view=3&white_background .)

Figure 6 shows the cosine-normalized comparison between these organizations. Without normalization Philips and Shell are distinguished as two separate groups at approximately the same positions on the map. This latter map (not shown here) can be web-started at http://www.vosviewer.com/vosviewer.php?map=http://www.leydesdorff.net/portfolio/aff_univ.map&network=http://www.leydesdorff.net/portfolio/aff_univ.net&label_size=1.40&n_links=1000&view=3&white_background .

Like factor analysis, cosine normalization enables the grouping into sets with communalities in the variance. In the case of portfolio analysis, however, one may wish to use co-occurrence matrices in order to observe the variance unique to the cases under study.

## 4. Diversity

Among the countries, Israel is indicated as the one with the greatest diversity in its portfolio in 2013; among the 20 cities the most diverse are Haifa, Beer Sheva, and Tel Aviv. From an evolutionary perspective, a diverse knowledge base can be expected to provide more opportunities for further knowledge development and related diversification (Heimeriks & Boschma, 2013).

**Table 3**: top-ten scores for countries, cities, and organizations in terms of diversity ($^2D^3$).

| Country | $^2D^3$ | N | Cities | $^2D^3$ | N | Organizations | $^2D^3$ | N |
|---|---|---|---|---|---|---|---|---|
| Israel | 1.4809 | 16,237 | Haifa | 1.4875 | 3,408 | Univ. of Amsterdam | 1.3805 | 6,040 |
| Spain | 1.4655 | 69,648 | Beer Sheva | 1.4574 | 1,905 | Philips | 1.3198 | 536 |
| UK | 1.4652 | 155,323 | Tel Aviv | 1.4551 | 4,206 | Samsung | 1.3173 | 1,494 |
| Germany | 1.4642 | 128,706 | Paris | 1.4518 | 24,877 | Georgia Inst. Technol. | 1.2743 | 3,523 |
| France | 1.4613 | 88,053 | Marseille | 1.4452 | 5,293 | Nestle | 1.2416 | 252 |
| Hungary | 1.4607 | 7,988 | Toulouse | 1.4375 | 5,899 | Pfizer | 1.2316 | 2,115 |
| Turkey | 1.4602 | 32,878 | Jerusalem | 1.4247 | 3,414 | LSE | 1.2049 | 1,170 |
| Luxembourg | 1.4561 | 1,073 | Shanghai | 1.4115 | 29,166 | Unilever | 1.2049 | 345 |
| Greece | 1.4543 | 13,533 | Atlanta | 1.3978 | 14,296 | Shell | 1.1279 | 179 |
| USA | 1.4540 | 553,620 | Eindhoven | 1.3963 | 2,554 | Google | 1.1153 | 198 |

Note that the University of Amsterdam is less diverse as an organization than Eindhoven as a city (in terms of journal portfolios). Of the 2,554 publications with Eindhoven as a city address, only 1,653 are consolidated in the database as from the "Eindhoven University of Technology". Other publications with an Eindhoven address are from medical research centers, hospitals, and startup companies. Note that one is allowed to make comparisons across units of analysis at different scales using $^2D^3$ for the measurement of interdisciplinarity. One can also compare units of analysis at different scales (e.g., a country and its universities).

**Conclusion**

In the vein of previous research efforts on portfolio mapping and analysis (e.g., Rafols *et al.*, 2010; Zhang *et al.*, 2009), we focused on portfolios in terms of the 10,000+ journals included in the Journal Citation Reports of WoS. The portfolios can be overlaid on the base map for these journals, but also—and perhaps more interestingly—they can be compared and analyzed statistically in terms of the differences among them. Using the matrix of 43 (leading) countries versus journals, we found a remarkably strong divide between advanced and less-developed nations. However, a more finely-grained analysis showed regional differences. Among both nations and cities, Israel scored highest on diversity in the portfolios. The differences among portfolios of universities when compared with relevant industries were significant.

At the methodological level, we noted that instruments that serve the grouping (such as cosine-normalization and factor analysis) can be counter-productive when one aims at visualizing the variation that is unique to each case. We also noted that the consolidated names in the database were not reliable in the case of using company names. The instrument, however, can be used with any document set retrieved from WoS, for example, for analyzing and comparing individual authors or document sets retrieved on the basis of informed search strings.

**Acknowledgement**
We are grateful to Thomson Reuters for JCR data.